\documentstyle[preprint,eqsecnum,aps]{revtex}
\begin{document}
\tightenlines
\preprint{UVA-INPP-97-06}
\title{\bf Differential cross sections for pion charge exchange \\
on the proton at 27.5 MeV}
\author{E.~Frle\v{z}, D.~Po\v{c}ani\'c, K.~A.~Assamagan,$^{*}$ 
        J.~P.~Chen,$^{\dag}$ K.~J.~Keeter,$^{\ddag}$ R.~M.~Marshall, 
        R.~C.~Minehart, and L.~C.~Smith}
\address{Department of Physics, University of Virginia, Charlottesville, 
Virginia 22901}
\author{G.~E.~Dodge,$^{\S}$ S.~S.~Hanna, and B.~H.~King}
\address{Department of Physics, Stanford University, Stanford, California 
94305}
\author{J.~N.~Knudson}
\address{Los Alamos National Laboratory, Los Alamos, New Mexico 87545}
\date{15 December 1997}
\maketitle
\begin{abstract}
We have measured pion single charge exchange differential cross sections 
on the proton at 27.5 MeV incident $\pi^-$\/ kinetic energy in the center 
of momentum angular range between $0^\circ$ and $55^\circ$. The extracted 
cross sections are compared with predictions of the standard pion-nucleon
partial wave analysis and found to be in excellent agreement.

\end{abstract}

\pacs{PACS Numbers: 13.75Gx, 25.80.Gn, 11.80Et}

\widetext
\tracingpages=1

\section{INTRODUCTION}
\label{sec:intro}

Absolute measurements of the pion-nucleon differential scattering cross 
sections below 150 MeV are sparse. The experimental database for the
pion single charge exchange reaction 
on the nucleon ($\pi N$\/ SCX) at these low energies is quite limited. 
The most recent pion-nucleon partial-wave analyses by 
Arndt~{\it et al.}~\cite{arn95,arn91} reflect that paucity of data.
Their parameterization neglects the expected isospin breaking effects and 
the interesting physics that could lie beyond the hadronic mass 
differences and the Coulomb interaction~\cite{gib95}. The data for 
the SCX reaction on free nucleons are, in addition, essential for our
understanding of nuclear medium effects on the $\pi N$ 
interaction~\cite{sig86}, such as multiple scattering processes and 
valence nucleon densities.

The question of the magnitude of the ``sigma term'' matrix element is also
not yet settled. The $\pi N$\/ $\sigma$-term explicitly breaks 
chiral symmetry in the effective QCD Lagrangian. Extrapolation of 
$\bar{D}^+$, an isospin-even $\pi N$ scattering amplitude, to the 
non-physical region leads to a value that is significantly larger than 
that extracted from the baryon mass spectra. The difference has been 
attributed to a nonzero $\bar{s}s$ quark content of 
the nucleon~\cite{gas82}. Inconsistencies between different 
$\pi N$ scattering experiments have impeded an unambiguous resolution 
of that discrepancy for a long time~\cite{koch82,gas91}.

The principal experimental difficulties in measuring the $\pi N$\/ SCX 
process below 150 MeV arise from the need for an accurate 
determination of the beam composition, absolute beam flux normalization, 
and accurate calibration of the $\pi^0$\/ detection efficiency. 
Two techniques have been used in the past to measure the $\pi p$ SCX 
differential cross sections. Early experiments~\cite{salom84,bagh88}  used
NaI crystal counters to detect a single photon from the final state 
$\pi^0$\/ decay. These measurements were performed at eight incident 
$\pi^-$\/ energies between 26.4 and 121.9 MeV and covered the laboratory 
polar angles between 0$^\circ$ and 145$^\circ$. The experimental 
uncertainties ranged from over 150\% at low energies and forward angles, 
to about 15\% at energies above 40 MeV and scattering angles larger than 
60$^\circ$.

Other published data~\cite{fitzg86} come from a study 
that used the LAMPF $\pi^0$\/ spectrometer for the coincident detection 
of two $\pi^0$\/ photons. This measurement was made at seven beam energies 
between 32.5 and 63.5 MeV and was restricted to laboratory polar 
angles smaller than 30$^\circ$. In a later LAMPF experiment 
Sadler~{\it et al.} used an electrostatic separator 
to obtain a pure pion beam in the energy range 10--40 MeV covering 
a selection of forward and backward center-of-momentum scattering angles;
their preliminary results are reported in Ref.~\cite{sad91}. 

Details of our experimental technique are given below. In 
Section~\ref{sec:beam} we discuss the critical issue of the $\pi^-$ beam
contamination which is large at low energies,
and show how we extracted the electron and muon beam fractions from
the measurements. Section~\ref{sec:targets} specifies the composition, 
dimensions
and geometry of the targets and the effective beam energies on targets. 
The integrated efficiency of the $\pi^0$ detector, discussed in
Section~\ref{sec:pi0accept}, is broken into several factors whose
values are determined in both calibration measurements and in a
Monte Carlo simulation.  The experimental cross sections are presented in
Section~\ref{sec:results} where they are compared  with the partial-wave
analysis prediction and previously published data.

\section{BEAM COMPOSITION AND FLUX NORMALIZATION}\label{sec:beam}

The measurements were performed in the Low-Energy Pion (LEP) channel at 
the Clinton P. Anderson Meson Physics Facility (LAMPF)~\cite{Bur75}. 
A weakly focusing 30 MeV $\pi^-$\/ beam tune was developed with 12 mr 
horizontal and vertical divergences, near-circular beam spot with a
diameter of 9 mm FWHM at target location, momentum  
spread $\Delta p/p$\/ of 3\%, and pion flux averaging 6$\cdot$10$^5$\ 
$\pi^-$/sec.

Relative on-target beam intensity was monitored with a gas ion chamber
in combination with a precision charge integrator. The chamber was a sealed
30 cm long aluminum cylinder with 125 $\mu$m steel windows, filled
with 0.0427 g/cm$^2$ of Argon gas.

Absolute cross-calibration of chamber ionization counts against the number
of pions in the beam was obtained through activation measurements of the
$^{12}$C($\pi^-$,$\pi N$)$^{11}$C reaction using $\o$70 mm $\times$ 3.2 mm
disk-shaped plastic scintillator targets (PILOT B scintillator, 91.6\%
$^{12}$C by weight)~\cite{But82}. The $^{11}$C activity of these targets,
measured after exposures to the $\pi^-$\/ beam of typically 20 minute
duration (one half-life of $^{11}$C) was well above background counting
rates. The background rates in the irradiated disks were constrained
separately in the analysis of each activation by independently measured
$e^+$-$\gamma$\/ detection efficiencies of the $^{11}$C counting
apparatus. The positron, photon and the coincident $e^+$-$\gamma$ signal
were on average six, three and one hundred times the background levels,
respectively. Polaroid films irradiated during the beam activations showed
ellipsoidal beam spots with major axes $\Delta x$$\times$$\Delta
y$=5$\times$3 cm$^2$ fully contained within the activation disk areas. The
focused beam pions, muons and electrons coming from the production target
produced over-exposed beam spots, while the muons from pions decaying in
flight left only a weak 10 cm diameter halo on Polaroid films placed at the
target position. The statistical reproducibility of the method outlined
above was better than 2.0\%.

The absolute calibration of the ion chamber was performed using higher
momentum $\pi^+$'s during our study of the
$\pi^+p$$\rightarrow$$\pi^0\pi^+p$\/ process near threshold~\cite{poc94},
i.e., the ratio of ionization counts $I_{\rm c}$ to the electron-equivalent
energy $\Delta E_{\rm ee}$ deposited in the chamber gas was determined
using ionizing particles in 160--260 MeV $\pi^+$\/ beams. Energy
depositions of different charged particles were calculated using the
Bethe-Bloch formula with appropriate corrections~\cite{leo87}.  Protons in
the $\pi^+$\/ beams were suppressed by means of a thin degrader in the beam
line. The residual proton fraction $f_p$=$N_p$/$N_{\pi^+}$\/ was deduced by
two independent methods: (i) a coincident $pp$\/ scattering using a liquid
$\rm H_2$ target of known thickness in conjunction with the published
differential $pp$\/ cross sections~\cite{arn95}, and (ii) $\pi^+p$\/
momentum separation scans. In the momentum scan measurements the ionization
count rate was determined as a function of the magnetic fields of the
channel dipole magnets and quadrupoles downstream of the thin degrader,
covering a 10\% range around the nominal momentum setting and thus allowing
easy mapping of the proton and pion beam momentum profiles.  The proton
contaminations $f_p$ of the $\pi^+$\/ beam tunes used for the ion chamber
calibration were found to be stable at 0.6 $\pm$ 0.1\%.  The ratio $I_{\rm
c}$/$\Delta E_{\rm ee}$ characterizing the ion chamber was established to
be (2.57 $\pm$ 0.11)$\cdot$$10^{-5}$ counts/MeV$_{\rm ee}$.

The activation measurements were affected by non-pionic contaminations of
the beam, i.e., electrons and muons.  The apparent number of $\pi^-$'s
deduced from the activation measurements using the 96 MeV/c incident beam, 
with the electron contamination $f_{e^-}$=$N_{e^-}$/$N_{\pi^-}$\/ and the muon
contamination $f_{\mu^-}$=$N_{\mu^-}$/$N_{\pi^-}$, had to be reduced by the
factor \begin{equation} 1+f_{e^-}{ {\sigma_{e^-}}\over {\sigma_{\pi^-}} }+
f_{\mu^-} { {\sigma_{\mu^-}} \over {\sigma_{\pi^-}} }, \end{equation} where
$\sigma_{e^-}$, $\sigma_{\mu^-}$\/ and $\sigma_{\pi^-}$\/ are the $e^-$,
$\mu^-$\/ and $\pi^-$\/ $^{11}$C activation cross sections, respectively.
The unpublished recommended low energy $e^-$ and $\pi^\pm$\/ activation
cross sections~\cite{leitch91} used in our analysis are listed in
Table~\ref{act} for the 30, 40, and 50 MeV incident $\pi^\pm$ beams.  The
cross section for the $^{12}$C($\mu^+$,$\mu^+n$)$^{11}$C reaction with a 60
MeV $\mu^+$\/ beam is known to be 21 $\pm$ 4 $\mu$b~\cite{kul72,ort78}.
Since $\mu^\pm$ activation cross sections are essentially
charge-independent we have used this datum in the absence of a $\mu^-$
measurement.  The incident particle threshold energy below which the
$^{11}$C activation cross section has to be zero is 18.7 MeV.  Using all of
the above data we obtained interpolated values of $^{11}$C activation cross
sections at energies appropriate for our activation measurements.  These
values, used in our analysis, are also given in Table~\ref{act}.

The ion chamber scaler counting rate was proportional to the factor 
\begin{equation}
N_{\pi^-}(\Delta E_{\pi^-}+f_{e^-}\Delta E_{e^-}+
f_{\mu^-}\Delta E_{\mu^-}),
\end{equation}
where the $\Delta E_{\pi^-}$, $\Delta E_{e^-}$\/ and $\Delta E_{\mu^-}$\/
corresponded to the $\pi^-$, $e^-$\/ and $\mu^-$ energy losses in the ion 
chamber gas. The activation measurements and ion chamber scaler counts were
used simultaneously to deduce the beam electron fraction
$f_{e^-}$. The value of the beam electron fraction, calculated by
interpolation from the LAMPF User's Handbook Table~6A-VII~\cite{How87}
for the 27.5 MeV LEP $\pi^-$\/ beam with the detectors located 2.5 m from
the channel exit quadrupole, was $N_{e^-}/N_{\pi^-}$=8.7. 
Fixing the $\mu^-$ fraction at $N_{\mu^-}/N_{\pi^-}$=0.75, consistent with 
on-line observations, gave the measured $f_{e^-}$ of 8.7 $\pm$ 1.5 (the quoted 
error is the standard deviation deduced from six independent activations). 
Varying the $\mu^-$ fraction between the outer limits of $f_{\mu^-}$=0.5 
and 1.5 changes the overall beam flux normalization only weakly, by 
$\sim$3.3\%. An example of the pion flux analysis for one representative
activation data set is shown in Fig.~\ref{contamination}.

The LEP channel is characterized by a background neutron flux of about
5$\cdot$10$^{-4}n$/$\pi^-$. The contribution of this background to $^{11}$C
activation via the reaction $^{12}$C($n$,$2n$)$^{11}$C was estimated to be 
$\le$0.5\% in Ref.~\cite{But82}.

The total correction to the number of ``observed $\pi^-$'s'', due to 
non-pionic contamination, amounted to 9.3\%. The overall systematic
uncertainty of the $\pi^-$\/ flux normalization was 7.4\% 
which reflects uncertainties in the $^{11}$C activation cross sections,
reproducibility and systematic uncertainties of our activation measurement 
method, as well as uncertainties of the electron and muon beam fractions.

\section{TARGETS}\label{sec:targets}

The LAMPF $\pi^0$\/ spectrometer~\cite{baer81} was used to detect $\pi^0$'s
produced in single charge exchange reactions on solid $\rm CH_2$ and
$\rm ^{12}C$ targets, as well as on three different, thin-walled liquid 
hydrogen (LH$_2$) targets.  The schematic layout of the experimental area is 
shown in Fig.~\ref{floor}.  The target properties are summarized in 
Table~\ref{targ}.  Density nonuniformities of the solid targets were 
determined to be $\le$1\%.

Solid targets were mounted in air without use of an evacuated scattering
chamber. They were oriented perpendicular to the beam direction with 
upstream faces positioned at the $\pi^0$\/ spectrometer pivot point.
This geometry improved the $\pi^0$\/ energy resolution due to the
partial compensation of the $\pi^0$ vertex uncertainty by the beam pion 
energy loss in the target.

The liquid $\rm H_2$ target cell and the scattering chamber surrounding it
were designed relying on insights gained in the analysis of data collected 
from two different, cylindrical Mylar target cells during initial test runs. 
The final target cell was a spherical copper flask with uniform wall
thickness of 5.0  $\pm$ 1.3 $\mu$m.  Fully filled with liquid H$_2$ the cell 
presented 142.4  $\pm$ 4.4 mb$^{-1}$ of hydrogen to the incident pion beam.

The LH$_2$ scattering chamber was shaped in the form of a drum with an
outer diameter of 55.9 cm and a 50.8 cm long horizontal axis aligned
perpendicular to the beam direction.  The cylindrical wall of the drum was
made of 1.3 cm thick aluminum.  Windows for beam entry and exit were 
cut in the walls of the cylinder and covered with a 25 $\mu$m thick Mylar
band wrapped completely around the cylinder to preserve vacuum tightness. 
All detected photon pairs originating from low energy $\pi^0$\/ decays in
the target region ($T_{\pi^0} \le 100$ MeV) passed through the chamber end
plate windows.  Each window consisted of a 13 $\mu$m thick Mylar sheet
sandwiched between two Kevlar layers of the same thickness.  On average,
the window matter traversed by each photon was equivalent to 0.013
radiation lengths.

The mounted targets were surveyed with a transit theodolite and their position 
at the spectrometer pivot point was always confirmed independently to within 
$\pm$1 mm by $^1$H$(p,p)p$\/ scattering measurements. The incident proton
beam spot was moved across the target in both the horizontal and vertical 
directions by varying the beam line bending magnet field values. 

The effective thicknesses of the target cells presented to the beam
particles and to the outgoing $\pi^0$\/ photons were calculated in a {\tt
GEANT} Monte Carlo simulation~\cite{brun87}, Table~\ref{targ}.  These
derived thicknesses were corrected subsequently for the fraction of the
$\pi^0\to\gamma\gamma$\/ photons converting in the target material,
scattering chamber, and spectrometer pre-radiators.  $f\rm{_a}$, the
probability for absorption of either $\pi^0$ decay photon in
material preceding the spectrometer converters, was calculated in the Monte
Carlo simulation and depended on the selected experimental geometry and the
target type; its values ranged from 12\% to 36\%.  $\Delta f_{\rm a}$, the
combined statistical and systematic uncertainty of the $f_{\rm a}$\/
values, was less than 1\%.

The average incident $\pi^-$\/ kinetic energies, integrated along the
thickness of the CH$_2$ or liquid H$_2$ target and weighted with the
$\pi^-$\/ beam profiles and beam energy straggling, were 27.5 $\pm$ 0.2 MeV
and 26.4 $\pm$ 0.2 MeV, respectively.  The absolute value of the beam
central momentum in the LEP channel is known with a 0.5\% accuracy.  That
uncertainty limit was set by measuring the energies of spallation particles
created at the pion production target~\cite{hoe82}. Beam momentum
reproducibility was better than $10^{-4}$ owing to the uncertainties in NMR
measurements of magnetic fields in the beam line bending magnets.


The upstream vacuum window of the scattering chamber was placed 16.5 cm 
upstream of the target. Consequently, the LH$_2$ target location was 11.5 
cm upstream of the scattering chamber center. This design reduced 
the background rates from SCX events in air by 40\%.  Extensive shielding
of an upstream section of the LEP beam line suppressed accidental
background rates in the $\pi^0$ detector to $\le$2\% of the $\pi^0$\/
signal (Fig.~\ref{tdcor}).

\section{\boldmath $\pi^0$\/ DETECTION: ACCEPTANCE AND EFFICIENCY}
\label{sec:pi0accept}

The data were taken with the $\pi^0$\/ spectrometer in the ``two-post''
configuration~\cite{gilad79}, with two arms (labeled J and K, respectively)
positioned symmetrically left and right with respect to the incident
$\pi^-$\/ beam (Fig~\ref{floor}).  Two different scattering polar angles
were selected for data taking, $0^\circ$ and $20^\circ$, with a $118^\circ$
opening angle between the two spectrometer arms and the 55 cm nominal
detector-to-target distance.  This configuration is optimized for maximum
geometrical acceptance of two coincident photons following the decay of a
25 MeV $\pi^0$.  The spectrometer's multiwire proportional chambers' (MWPC)
fiducial cuts imposed in the analysis required a reconstructed photon
conversion vertex in the spectrometer lead glass detectors to lie within a
pyramidal volume whose apex coincides with the target center and whose base
is a rectangle located two radiation lengths deep in the spectrometer
calorimeter blocks extending to the calorimeter edges.  The detector
acceptance for monoenergetic $\pi^0$'s with the nominal incident kinetic
energy of 30 MeV was calculated with the Monte Carlo program {\tt
PIANG}~\cite{gilad79} and the results are listed in Table~\ref{exp}.
Essentially the same effective solid angle values were obtained in a {\tt
GEANT} model of the detector response. Comparing the {\tt PIANG} and {\tt
GEANT} calculations leads to an estimated 3\% systematic uncertainty of the
$\pi^0$ angular acceptance. The uncertainty is dominated by electromagnetic
losses near the margins of the fiducial areas.  The energy lineshapes of
the detected $\pi^0$'s from CH$_2$ and liquid H$_2$ targets are compared to
the simulated $\pi^0$\/ energy spectra in the two panels of
Fig.~\ref{energy}.

The $\pi^0$\/ spectrometer detection efficiency is an important factor in
determining the overall uncertainty of the cross sections because of the 
complexity of the instrument. In the past, the spectrometer instrumental 
efficiency was calibrated to about 1\% accuracy at the $\pi^-$ beam momentum 
of 522 MeV/c~\cite{gaille84}, but it is significantly less well understood
at the low momenta used in the present work.

We used penetrating cosmic muons to measure directly the intrinsic 
instrumental efficiencies of the lead glass detectors, plastic scintillator 
elements, and the multiwire proportional chambers. J and K arm 
calibrations were performed independently, collecting $\ge$$10^5$ cosmic muon
events in each arm, resulting in approximately a 0.3\% statistical 
uncertainty in the deduced detector efficiencies.

The spectrometer $\pi^0$\/ detection efficiency $\epsilon_{\pi^0}$\/
can be decomposed into a product of individual efficiencies~\cite{frlez93}:
\begin{equation}
\epsilon_{\pi^0}=\epsilon^{\rm JK}_{\pi^0}\epsilon_{\rm m}\epsilon_{\rm c}
\epsilon_{\rm s}\epsilon_{\rm t}\epsilon_{\rm b},
\end{equation}
where $\epsilon_{\pi^0}^{\rm JK}$\/ is the simultaneous conversion 
probability for both
$\pi^0\to\gamma\gamma$ decay photons in the J and K arms, $\epsilon_{\rm
m}$ is the weighted combined wire chamber efficiency, $\epsilon_{\rm c}$ is
the converter ``transparency'' for the charged showers (defined below),
$\epsilon_{\rm s}$ is the weighted scintillator efficiency for minimum
ionizing particles (MIPs), $\epsilon_{\rm t}$ represents the tracking
algorithm efficiency for the accepted photon showers, and $\epsilon_{\rm
b}$ is a small correction due to the shower back-splash. Some of these
efficiencies depend weakly on the $\pi^0$\/ kinetic energy and direction,
the $\gamma$-$\gamma$\/ energy asymmetry and the photon conversion point
positions.  These dependencies were studied and have been taken into
account in the {\tt GEANT} simulations of the spectrometer response
described in the following text.

The two-photon conversion probability $\epsilon_{\pi^0}^{\rm JK}$\/ is 
a function of the single converter plane conversion probability 
$\epsilon_\gamma$:
\begin{equation}
\epsilon_{\pi^0}^{\rm JK}=[1-(1-\epsilon_\gamma)^3]^2.
\end{equation}
The quantity $\epsilon_\gamma$\/ was extracted directly from our data in 
an off-line analysis of all
recorded $\pi p$\/ SCX events. The $\pi^0$\/ events at the $\pi^-$\/ 
incident energy of 27.5 MeV involved detection of coincident photon pairs 
with each $\gamma$ having the energy of $E_\gamma \simeq 82$ MeV. 
The distribution of 
triggered conversion plane pairs was tabulated in a 3$\times$3 matrix.
Each entry in the matrix corresponded to the number of good photon
conversions in a given (J,K) pair of arm converter planes.  The efficiency
$\epsilon_\gamma$\/ was then calculated in a simultaneous fit to all nine
matrix elements. 

The value $\epsilon_\gamma$\/ has also been previously determined
semi-empirically by the equation~\cite{baer81}:
\begin{equation}
\epsilon_\gamma=0.86[0.327+0.1\log(0.01E_\gamma\hbox{(MeV)}],
\end{equation}
with parameters based on photon interaction probabilities~\cite{hubb69} and
the known converter specifications. The above relation, Eq.~(4.3), places 
$\epsilon_\gamma$\/ just 2.5 standard deviations below our measured value 
of 0.292 $\pm$ 0.006. 

The analyzed event fraction $\eta_{\rm a}$, defined as the ratio of the number
of $\pi^0$\/ hardware triggers to the number of ``analyzable'' events 
with good wire chamber information, was understood entirely in terms of the
instrumental MWPC efficiencies. Over the period of the experiment, for each
individual run, $\eta_{\rm a}$ was equal (within the associated statistical 
uncertainty) to the appropriately weighted product of six intrinsic wire 
chamber efficiencies.
The average efficiency $\epsilon_{\rm m}$\/ of a single MWPC chamber varied 
between 94.4\% and 95.6\% over one month of data collection. The average 
veto counter and scintillator counter efficiencies, appropriately weighted 
with photon conversion probabilities in three conversion planes, 
were calculated to be 96.9\% and 96.1\%, respectively.

The efficiency of the electromagnetic shower tracking algorithm was extracted 
independently from the 27.5 MeV SCX runs with LH$_2$ and CH$_2$ targets, 
after subtraction of the appropriate target-empty and $^{12}$C target 
backgrounds from the data. The ratio of
the number of events which survived all analyzer cuts to the number of 
events that satisfied less restrictive conditions for good MWPC hits inside
predefined fiducial areas, was defined as the tracking efficiency
$\epsilon_{\rm t}$. The measured tracking efficiency was stable for all
collected data sets and averaged 0.76 $\pm$ 0.02, as demonstrated in
Fig.~\ref{track}.

All relevant instrumental and software aspects of $\pi^0$\/ detection in 
the spectrometer were studied in a complementary way and in greater detail 
in a {\tt GEANT} simulation~\cite{brun87} in order to provide a cross-check 
for the measured efficiencies. In this simulation the required material 
properties of Schott LF5 lead glass (of which the converters and total 
absorption blocks are made), were taken from the original manufacturer's 
specification~\cite{baer80}.  The Monte Carlo calculation yielded 
a 29.2 $\pm$ 2.0\% single-plane conversion efficiency.  An event was
counted as a ``good'' $\pi^0$ $\gamma$ conversion if a photon, interacting
in the converter material by the photoelectric effect, Compton scattering,
or pair production, generated secondary particles that deposited more than
50 MeV in the lead glass calorimeter.  The same energy threshold for a
single spectrometer arm was used in the data analysis.  The agreement
between measured and simulated probabilities quoted above confirms that we 
had specified the appropriate converter composition and analysis cuts.

Simulated showers that converted into neutral events inside the converter or 
events that failed to provide the necessary tracking pulses in scintillators 
and wire chamber planes had to be taken into account separately. The 
probability that a photon from $\pi^0$ decay generates a shower with at least 
one detectable charged particle in the volume occupied by the MWPC planes 
defined the converter transparency $\epsilon_{\rm c}$\/. In high-statistics 
simulations, opacity, i.e. the inefficiency of shower tagging 
$(1-\epsilon_{\rm c})$, converged to a value of 5.6 $\pm$ 0.2\%. This
probability should be compared with the previous converter opacity 
measurement of 7.2 $\pm$ 0.5\% for 15\% thicker converters~\cite{gilad77}.

Particularly important for the determination of tracking efficiency were 
tests imposed in the analyzer {\tt TRACER} routine which reconstructed the 
trajectories of the charged shower particles through 
a spectrometer arm. 
The {\tt TRACER} tracking reconstruction was simulated in the {\tt GEANT 
GUSTEP} subroutine where shower particles were tracked through all elements
of the experimental apparatus. Efficiency parameters, relating the
response of the wire chambers to the minimum ionizing particles, were 
inferred from the high-statistics cosmic muon measurements. 
The following cuts were implemented:

\begin{itemize}
\item The photon shower coordinates were reconstructed independently
from (a) the MWPC wires that were hit, and (b) from the shower energy
distribution inside the calorimeter lead glass blocks.  Both sets of
coordinates were projected back to the scintillator plane immediately
following the conversion plane where the shower originated.  The two
projected points were required to fall inside an acceptance window of
$\Delta x$$\times$$\Delta y$=10$\times$20 cm.
\item After photon conversion the electromagnetic shower is tracked
through at least two wire chamber planes. On the basis of the resulting
MWPC wire hits a shower track direction was reconstructed.
This direction was compared with the line connecting the target center
and the conversion point (defined above). The relative angle between
the two lines was required to be $\le$18$^\circ$.
\item If in the tracking of a shower any individual MWPC plane reported 
more than four wires hit, the event was discarded.
\end{itemize}
These cuts were identical to the ones imposed in the data analysis.  The
tracking efficiency $\epsilon_{\rm t}$ deduced from the percentage of
simulated $\pi^0\to\gamma\gamma$ photon conversions surviving all  
analyzer cuts was 0.73 $\pm$ 0.03, where most of the uncertainty is due to 
approximations involved in the Monte Carlo description of the MWPC geometry 
and response.  The Monte Carlo value is in good agreement with the
experimental value, $\epsilon_{\rm t}$=0.76 $\pm$ 0.02, listed above.  The
latter value was used in the calculation of the differential cross sections
(see also Figs.~\ref{track}~and~\ref{geant}).

In summary, the detection efficiency of the $\pi^0$\/ spectrometer was 
calibrated for the photon energy range 50--100 MeV with a 4.6\% uncertainty. 
The ingredients entering the detection efficiency calculation are summarized 
in Table~\ref{eff}. The integrated value of $\epsilon_{\pi^0}$\/ for our 
spectrometer settings and choice of adjustable analyzer cuts was 
0.175 $\pm$ 0.008. The general approach outlined in this section, 
however, can be followed to calculate or measure the spectrometer detection 
efficiency for any chosen set of applied tests as well as for different 
$\pi^0$\/ energies.

\section{RESULTS AND DISCUSSION}\label{sec:results}

Collecting the results presented in
Sections~\ref{sec:beam}--{\ref{sec:pi0accept}, single charge exchange
differential cross sections were evaluated for six $8^\circ$-wide polar
angle bins. The acceptable angular bin sizes were restricted by the
5.5$^\circ$ $\rm rms$\/ directional resolution of the spectrometer and by
the event statistics. The variation of the cross section over the range of
pion energies present in the target (due to the combination of nonzero
$\pi^+$ beam momentum spread and $\pi^+$ energy loss in the target) was
taken into account by assuming the cross section energy dependence of the
VPI SM95 partial-wave solution~\cite{arn95}.  This cross section scaling
correction, $f_\sigma$, was bracketed within the ($-$0.5\%,$+$1.0\%) range
and all extracted cross sections were referred to the central beam
on-target energy of 27.5 MeV.

The differential cross sections were calculated using the expression:
\begin{equation}
{ {d\sigma}\over {\Delta\Omega} }(\theta_{\rm CM})={ { {Y}Jf_{\sigma} }\over
{ N_{\pi^-}t \Delta\Omega_{\pi^0} \epsilon_{\pi^0} (1-f_{\rm a}) \Gamma_{\pi^0
\rightarrow\gamma\gamma} \eta_{\rm c} \eta_{\rm v} } },
\end{equation}
where $Y$\/ is the number of detected $\pi^0$'s in a given angular bin after 
background subtraction, $J$\/ is the Jacobian of transformation from 
the laboratory (LAB) to the center-of-momentum (CM) frame, $f_{\sigma}$\/ is
the factor defined just above, $N_{\pi^-}$\/ is
the number of beam $\pi^-$'s incident on a target, $t$\/ is the effective
target thickness, $\Delta\Omega_{\pi^0}$\/ is the laboratory solid angle
of an angular bin, $\epsilon_{\pi^0}$\/ is the integrated $\pi^0$\/ detection 
efficiency for a given bin, $(1-f_{\rm a}$) is the fraction of photons not
absorbed before conversion, $\Gamma_{\pi^0\rightarrow
\gamma\gamma}$\/ is the $\pi^0$$\rightarrow$$\gamma\gamma$\/ decay branching 
ratio, $\eta_{\rm c}$\/ is the computer live time fraction, and 
$\eta_{\rm v}$\/ is the spectrometer veto live time fraction.

The experimental angular distribution is plotted in Fig.~\ref{diff} and
results are summarized in Table~\ref{exp}, together with the comparison
with the results of the latest pion-nucleon phase-shift analysis by the VPI
group SM95~\cite{arn95}.  Using the analysis presented here, our
experimental yields lead to differential cross sections that are 1.01 $\pm$
0.06 times the VPI SM95 partial-wave solution in the angular range covered,
$\theta$=0$^\circ$--55$^\circ$.  The overall normalization uncertainty of
the experiment 8.7\% is due to the 7.4\% uncertainty in the pion flux
(Section~\ref{sec:beam}) and the 4.6\% uncertainty of $\epsilon_{\pi^0}$,
the $\pi^0$ spectrometer detection efficiency
(Section~\ref{sec:pi0accept}).  Combining the statistical and systematic
uncertainties of the data points we find all our measurements fall within
one standard deviation of the partial-wave predictions.  While the $\pi N$
amplitude analysis has been plagued by inconsistencies in the experimental
data base below about 150 MeV since the 1980's, our results confirm the
validity of the ``standard'' pion-nucleon phase shift analysis in the
important low energy region.  Incorporation of our data points, which have
smaller statistical and systematic uncertainties then the previous
measurements, into the low energy SCX database will have a constraining
influence on future partial-wave analyses.

We note that the similar low energy measurement of Fitzgerald
{\it et al.}~\cite{fitzg86}, performed with the same instrument, should be
corrected with the new and more precise $^{11}$C activation cross
sections~\cite{leitch91}, listed in Table~\ref{act}.  Their appropriately 
renormalized differential cross sections at 32.5 MeV are then no longer 
one standard deviation above the VPI SM95 $\pi p$\/ SCX partial-wave 
solution, but are between 2.0 and 2.4 times the predicted values, or more 
than 4.5 standard deviations away.

We thank the staff of LAMPF groups MP-7 and MP-8 for strong technical 
support, without which our measurements would not have been possible. 
We are particularly indebted to Bob Garcia who was responsible for the 
liquid hydrogen target design and construction. We acknowledge  
fruitful discussions with M. Sadler. This work was supported
by the U.S. National Science Foundation and U.S. Department of Energy.

\begin{figure}
\caption
{The total number of beam $\pi^-$'s deduced from the number of ion
chamber counts (full line) and the activation measurement (dotted line) as
a function of the beam electron fraction $N_{e^-}/N_{\pi^-}$\/ for one
representative run. The $\mu^-$\/ contamination affects significantly only
the ion chamber counts. Using six independent activations and fixing the 
$N_{\mu^-}/N_{\pi^-}$\/ ratio at 0.75 (see text) we find 
the $e^-$ fraction of 8.7 $\pm$ 1.5, in good agreement with the LAMPF User 
Handbook~\protect\cite{How87}. The shaded bands demonstrate the
uncertainties of the $\pi^-$, $e^-$ and $\mu^-$ activation cross sections
and the accuracy with which the ionization losses were known.
}
\label{contamination}
\end{figure}

\begin{figure}
\caption
{Layout of the LEP experimental channel for the present experiment (top
view) showing the arrangement of the $\pi^0$\/ spectrometer arms for 
an opening angle of 118$^\circ$ that optimizes the acceptance of the 25 
MeV $\pi^0$'s, as well as the scattering chamber, ion chamber, beam pipe 
and shielding walls. The bottom panel is a schematic drawing of the
$\pi^0$\/ spectrometer from Ref.~\protect\cite{baer81}. The orientation 
of J and K arms in the two post configuration is shown. The details 
of the spectrometer arms, with three sets of converter, scintillator, 
and MWPC detectors, as well as the 3$\times$5 array of lead-glass total
absorption blocks can be seen. 
}
\label{floor}
\end{figure}

\begin{figure}
\caption{Histogram of relative timing between the two arms (J and K) of 
the LAMPF $\pi^0$\/ spectrometer for the $\pi^0$\/ SCX events on the 
$\rm CH_2$ target at 27.5 MeV. TDC values for the scintillator planes in J 
and K arms were corrected for the photon time-of-flight between the event 
target vertex and photon conversion points as well as for the light 
propagation delay in the scintillator planes. The achieved timing resolution
was 1.37 ns FWHM. The crucial feature is the absence of an accidental 
background: virtually all events ($\ge$98\%) in the histogram are real 
$\pi^0$'s.}
\label{tdcor}
\end{figure}

\begin{figure}
\caption{The net subtracted $\pi^0$\/ kinetic energy spectra
for the single charge exchange reaction on the proton at 27.5 MeV and
26.4 MeV, respectively.
Top panel shows the data obtained with the 0.711 $\rm {g/cm^2}$ thick 
$\rm {CH_2}$ target. The $\rm ^{12}C$ contribution was measured and corrected
for by measuring the $\pi^0$\/ yield from an equivalent-thickness carbon 
target. The bottom panel shows 
the spectrum acquired with the 0.236 $\rm {g/cm^2}$ liquid hydrogen target,
(LH$_2$ ``C'' in Table~\ref{targ}).
Solid histograms represent results of Monte Carlo calculations of
the $\pi^0$\/ spectrometer acceptance with the modified {\tt PIANG} 
code~\protect\cite{gilad79}. The $\gamma$-$\gamma$\/ energy asymmetry cut
$X$=$(E_{\rm J}-E_{\rm K})/(E_{\rm J}+E_{\rm K})$$\le$0.2 was applied 
to both measured and simulated events.}
\label{energy}
\end{figure}

\begin{figure}
\caption{The average tracking efficiency $\epsilon_{\rm t}$ calculated for 
all SCX runs was 0.76 $\pm$ 0.02. The data collected with 
the solid CH$_2$ target as well as with the liquid H$_2$ 
targets were included in the quoted average. The absence of accidental
backgrounds implies constant  $\epsilon_{\rm t}$, 
independent of the detector geometry and influenced only by the tracking 
cuts used in the analysis. The data points confirm that expectation.
}
\label{track}
\end{figure}

\begin{figure}
\caption{A {\tt GEANT} simulation of the tracking efficiency of the $\pi^0$\/ 
spectrometer analyzer code. Neutral pions generated from 27.5 MeV 
$\pi^-$\/ beam SCX interactions in the CH$_2$ target were identified by 
the electromagnetic showers tracked through 
the volume of the modeled spectrometer arm. On average, the showers 
produced 1.37 charged minimum-ionizing particles exiting
the converter. The percentage of the two-arm $\pi^0$ decay photon conversions
surviving the {\tt TRACER} window and slope cuts on the reconstructed 
trajectories and passing a limit on the maximum number of hit wires in this 
simulation was 73 $\pm$ 3\%. That result should be compared
with the measured tracking efficiency of 76$\pm2$\%. The panels show (a) 
measured (full histogram) and simulated (shaded histogram) energy spectra in
a lead glass calorimeter, (b) a distribution of the energy-weighted coordinates 
of the hit blocks in the segmented 3$\times$5 element lead 
glass calorimeter, (c) differences between the coordinates of 
a MWPC-reconstructed $\gamma$\/ conversion point and the energy-weighted 
lead block energy deposition location, and (d) measured and simulated 
``best'' angle between the back-projected line from the target center 
to the conversion point deduced from hits in X and X$'$ wire chambers 
and to a shower's center-of-gravity in lead-glass blocks. All histograms
(measured or simulated) shown in the four panels correspond to ``good''
$\pi^0$ events only.}
\label{geant}
\end{figure}

\begin{figure}
\caption{Measured differential cross sections for 
the $\pi^-p$$\rightarrow$$\pi^0n$\/ reaction at 27.5 $\pm$ 0.2 MeV. 
The plotted cross sections were obtained by subtracting the measured 
$\rm ^{12}C$ contribution from the $\pi^0$ yields with the $\rm CH_2$ 
target. The plotted error bars are statistical uncertainties only, calculated 
from numbers of detected events. In addition, an overall normalization 
uncertainty of 8.7\% applies to all points (see text).
The full line represents the VPI SM95 $\pi p$\/ SCX partial-wave 
solution~\protect\cite{arn95}.
}
\label{diff}
\end{figure}

\mediumtext
\begin{table}
\caption{Recommended $^{12}$C($\pi^-$,$\pi N$)$^{11}$C activation cross 
sections for 30, 40, and 50 MeV $\pi^\pm$\/ beams and cross sections
for $^{11}$C production by electrons of the same momenta from 
unpublished measurements of Leitch~{\it et al.}~\protect\cite{leitch91}.
The $e^-$ activation cross section $\sigma_{e^-}$\/ at 128 MeV/c 
was published by Kuhl and Kneissl~\protect\cite{kul72}. 
The $\mu^+$-induced $^{11}$C production measured with a 60 MeV
$\mu^+$ beam gave five hundred times smaller cross sections than the 
associated $\pi^+$ activity~\protect\cite{ort78}. Consequently, the 
activation cross sections weighted over the $\pi^-$\/ beam momentum 
spread in the activation target disk used in our analysis were 
$\sigma_{\pi^-}(\bar{T}_{\pi^-}$=28.7 MeV)=1.50 $\pm$ 0.07 mb,
$\sigma_{e^-}(\bar{T}_{e^-}$=94.6 MeV)=64.4 $\pm$ 3.4 $\mu$b, and
$\sigma_{\mu^-}(\bar{T}_{\mu^-}$=36.2 MeV)=9.1 $\pm$ 1.7 $\mu$b,
respectively.
The fifth column shows the ratio of the unpublished  $\pi^-$-induced 
activation cross sections from the LAMPF experiment E942 to the older 
values that were used for $\pi^-$ beam flux normalization in 
the $\pi p$\/ SCX experiment of~Ref.\protect\cite{fitzg86}. 
}
\bigskip
\label{act}
\begin{tabular}{cccccc}
$p_{\rm beam}$ & $T_{\pi^\pm}$ & $\sigma_{\pi^+}$ & $\sigma_{\pi^-}$ &
$\displaystyle {\sigma_{\pi^-}\ {\rm Ref.}~\protect\cite{leitch91} \over 
\sigma_{\pi^-}\ {\rm Ref.}~\protect\cite{drop79}} $ & $\sigma_{e^-}$ \\
(MeV/c) & (MeV) & (mb) & (mb) & \ & (mb) \\
\tableline
$17.3$&  18.7& $0.0$& $0.0$&  0.0& 0.0 \\
$96.3$&  30.0& $3.2\pm 0.4$& $1.70\pm 0.08$&  $1.89$& $0.0664\pm 0.0035$ \\
$113.0$& 40.0& $6.5\pm 0.4$& $3.89\pm 0.15$&  $1.34$& $0.0954\pm 0.0140$ \\
$128.3$& 50.0& $10.3\pm 0.6$& $6.10\pm 0.50$& $1.00$& $0.124 \pm 0.020$  \\
\end{tabular}
\end{table}

\begin{table}
\caption{List of targets used in the SCX measurements. For liquid hydrogen 
(LH$_2$) targets the geometrical diameters of cylindrical and 
spherical cells are quoted. Neutral pion yields measured with $^{12}$C 
targets in the $\pi^-$ beam were used to subtract the carbon contribution
in CH$_2$ target data.}
\label{targ}
\bigskip
\begin{tabular}{cccccc}
Target & Description &Thickness & Areal Density &Areal Density & Volume
                                                                Density \\ 
(Symbol) & \ & (mm) & (g/cm$^{2}$) & (mb$^{-1}$) & (g/cm$^{3}$) \\ 
\tableline
$^{12}$C ``A''& Graphite Sheet     & $3.18\pm 0.02$  & $0.5289\pm 0.0040$& 
 $26.54\pm 0.17$& 1.660  \\
$^{12}$C ``B'' & Graphite Sheet    & $6.82\pm 0.02$  & $1.0787\pm 0.0045$& 
 $54.13\pm 0.35$& 1.582  \\
$^{12}$C ``C'' & Graphite Sheet    & $3.40\pm 0.02$  & $0.5374\pm 0.0023$& 
 $26.97\pm 0.17$& 1.581  \\
$^{12}$C ``D'' & Graphite Sheet    & $4.95\pm 0.02$  & $0.7826\pm 0.0050$& 
 $39.27\pm 0.25$& 1.581  \\
CH$_2$         & Polyethylene Plate& $7.77\pm 0.01$  & $0.7112\pm 0.0020$& 
 $91.77\pm 0.26$& 0.920  \\
LH$_2$ ``A''  & Vert. Mylar Cyl.   & \o$38.1\pm 1.0$ & $0.247\pm 0.007$  & 
 $149.2\pm 4.6$ & 0.070  \\
LH$_2$ ``B''  & Horiz. Mylar Cyl.  & \o$38.1\pm 1.0$ & $0.214\pm 0.006$  & 
 $129.3\pm 4.0$ & 0.070  \\
LH$_2$ ``C''  & Copper Sph. Bulb   & \o$38.1\pm 1.0$ & $0.236\pm 0.006$  & 
 $142.4\pm 4.4$ & 0.070  \\
\end{tabular}
\end{table}

\begin{table}
\caption{Factors contributing to the total $\pi^0$\/ detection 
efficiency given by the integral value $\int\epsilon_{\pi^0} d\Omega_{\pi^0} 
dT_{\pi^0}$=0.175 $\pm$ 0.008. 
The measurement~\protect\cite{gilad79} is scaled down for the new thinner 
converters but corresponds to 100 MeV photons as compared to lower 
energy gammas from our Monte Carlo simulation ($\sim$82 MeV $\gamma$'s 
from 27.5 MeV $\pi^0$'s decays). The uncertainties listed in the fifth column 
are the combinations of statistical and estimated systematic (when 
applicable) standard deviations.}
\label{eff}
\bigskip
\begin{tabular}{cllcc}
Symbol & \multicolumn{1}{c}{Description} & \multicolumn{1}{c}{Method} 
& Efficiency &Error \\
  & & & (\%) & (\%) \\
\tableline
$\rm \epsilon_\gamma$\hfill& single converter detection efficiency\hfill
 &SCX $\pi^0$ detection~\protect\cite{frlez93}\hfill& 29.2& 2.0 \\
 & & $\gamma$ attenuation coefficients~\protect\cite{hubb69}\hfill& 27.9&
                                                             1.0 \\[2ex] 
$\rm \epsilon_m$\hfill& instrumental MWPC efficiency\hfill
 &cosmic muon trigger~\protect\cite{frlez93}\hfill& 96.1& 0.2    \\ [2ex]
$\rm \epsilon_c$\hfill& converter transparency for\hfill
 &{\tt GEANT} simulation~\protect\cite{frlez93}\hfill& 88.9& 0.4 \\
 & minimum ionizing particles\hfill& experiment~\protect\cite{gilad79} 
                                            \hfill& 87.6& 1.0    \\[2ex]
$\rm \epsilon_s$\hfill& weighted scintillator efficiency\hfill
 &cosmic muon trigger~\protect\cite{frlez93}\hfill& 96.2& 0.5    \\[2ex]
$\rm \epsilon_p$\hfill& maximum number of\hfill
 &cosmic ray trigger+SCX~\protect\cite{frlez93}\hfill& 92.4& 1.0\cr
 & charged particle prongs\hfill& tagged $\gamma$ beam\
                        \protect\cite{gilad79}\hfill&  91.4& 2.0 \\[2ex]
$\rm \epsilon_d$\hfill& {\tt TRACER} shower window cuts\hfil
 &{\tt GEANT} simulation \protect\cite{frlez93}\hfill& 73.0& 3.0 \\
 & &SCX $\pi^0$ detection~\protect\cite{frlez93}\hfill& 76.0& 2.0 \\[2ex]
$\rm \epsilon_v$\hfill& weighted veto efficiency\hfill
 &cosmic muon trigger~\protect\cite{frlez93}\hfill& 96.9& 0.5     \\[2ex]
$\rm \epsilon_b$\hfill& backsplash self-vetoing\hfill
 &{\tt GEANT} code \protect\cite{frlez93}\hfill& 99.4& 0.2        \\
\end{tabular}
\end{table}

\narrowtext
\begin{table}
\caption{Experimental differential cross sections for the $\pi p$ SCX 
reaction at 27.5 $\pm$ 0.2 MeV, measured using the $\rm CH_2$ target with 
hydrogen thickness of 0.0612 $\rm g/cm^{2}$. The comparison with the VPI
partial-wave analysis SM95~\protect\cite{arn95} is shown in the last
column. The quoted error bars are statistical uncertainties from the
measured yields, background count subtractions, and Monte Carlo acceptance
statistics. There is an additional overall 8.7\% systematic uncertainty,
due to the pion flux normalization and $\pi^0$ spectrometer detection
efficiency (see text). It applies to all six measured cross sections.}
\label{exp}
\bigskip
\begin{tabular}{ccccc}
$\langle \cos\theta_{\rm CM}\rangle$ & Yield & $\Delta\Omega_{\rm CM}$
& ${d\sigma/d\Omega}\vert _{\rm CM}$ & $\displaystyle{
{ {{d\sigma}/{d\Omega}\vert ^{\rm E1179}_{\rm CM}} }\over 
{d\sigma/d\Omega\vert ^{\rm SM95}_{\rm CM} } }$  \\
\ & $Y$ & (msr) & ($\mu$b/sr) & \hbox{\ \ \ } \\
\tableline
0.99664&  37.3 $\pm$  8.4&  2.338 $\pm$ 0.011&  $59\pm13$& 0.99 $\pm$ 0.23 \\
0.96998& 129.2 $\pm$ 14.5&  6.994 $\pm$ 0.020& $68\pm7$& 1.04 $\pm$ 0.12 \\
0.91769& 254.2 $\pm$ 19.6& 10.967 $\pm$ 0.024& $86\pm7$& 1.12 $\pm$ 0.09 \\
0.84173& 311.7 $\pm$ 21.5& 12.163 $\pm$ 0.026& $95\pm7$& 1.00 $\pm$ 0.07 \\
0.74494& 304.5 $\pm$ 20.5&  9.644 $\pm$ 0.023& $120\pm7$&1.00 $\pm$ 0.06 \\
0.63070& 211.5 $\pm$ 15.4&  5.772 $\pm$ 0.017& $136\pm9$& 0.91 $\pm$ 0.06 \\
\end{tabular}
\end{table}

\input psfig.sty
\centerline{\psfig{figure=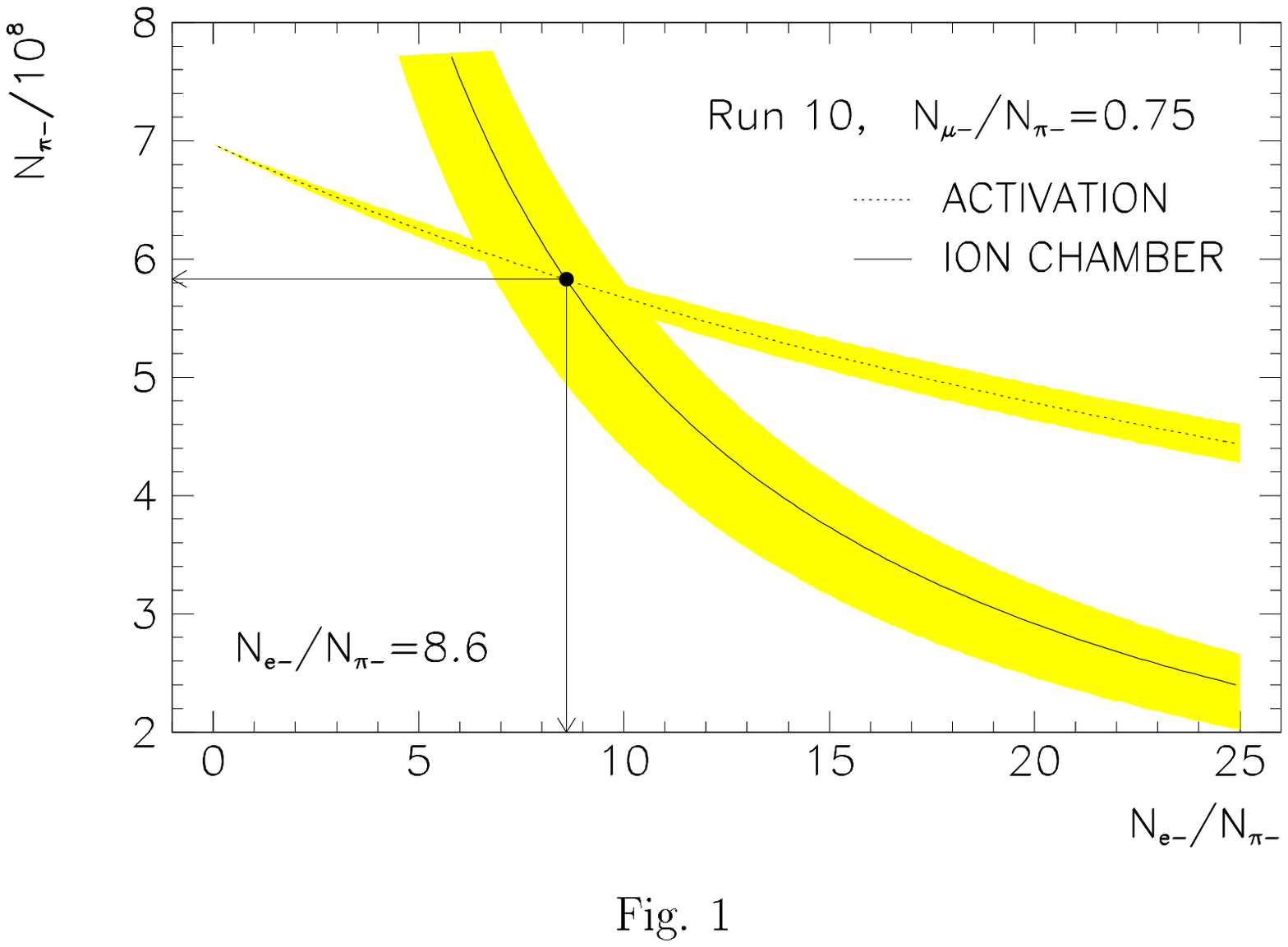}}
\newpage
\centerline{\psfig{figure=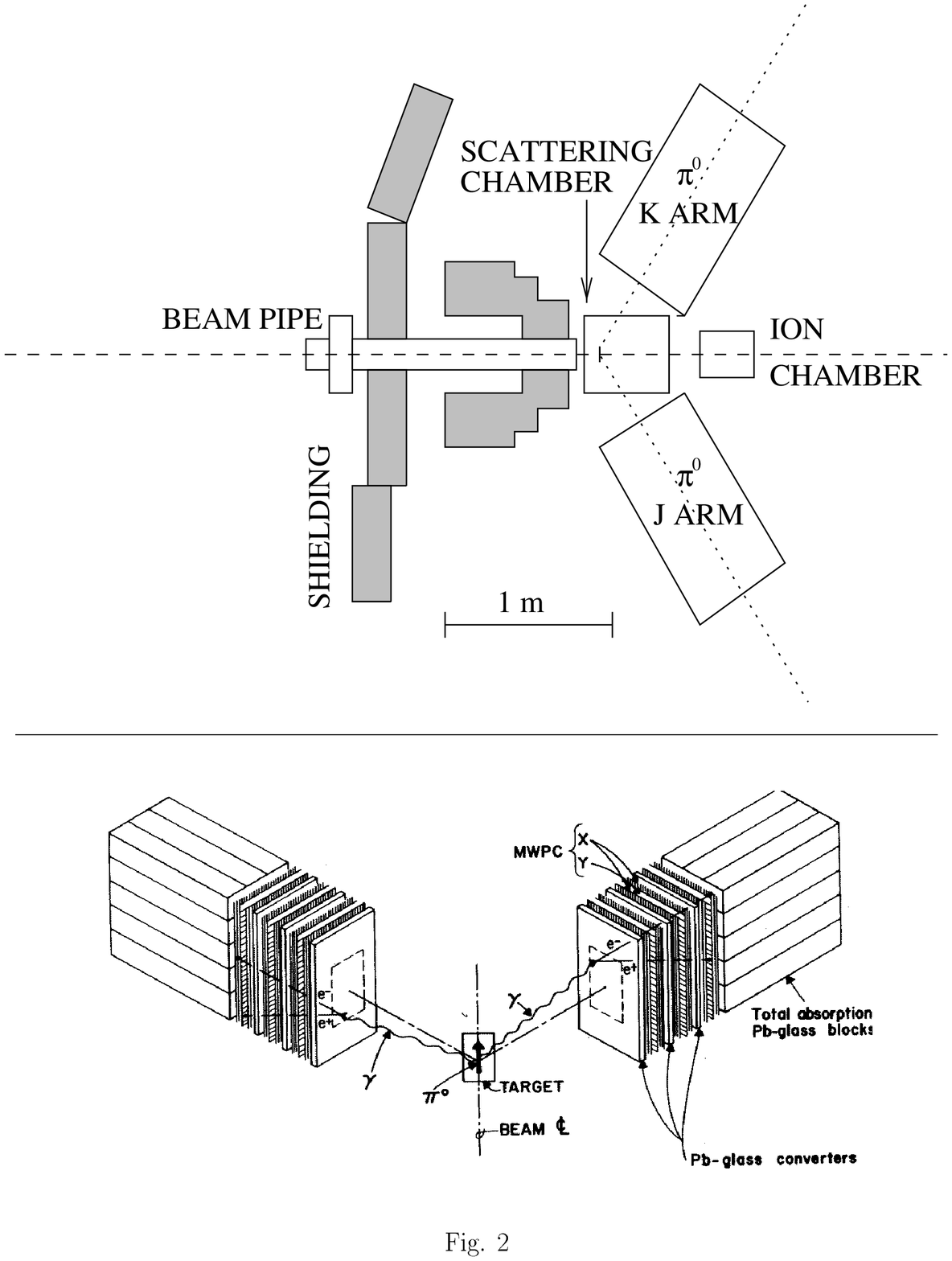}}
\newpage
\centerline{\psfig{figure=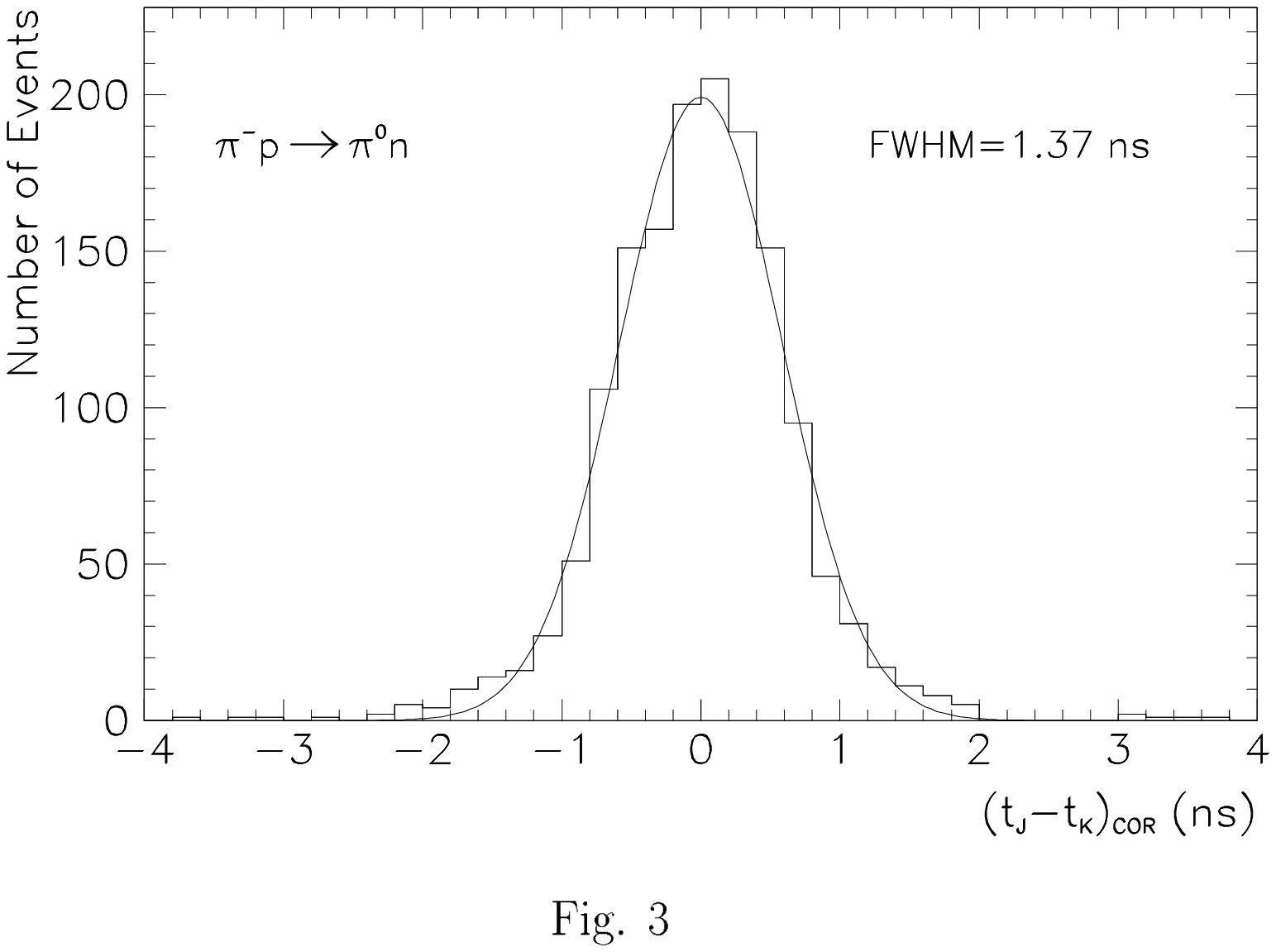}}
\newpage
\vglue -3cm
\centerline{\psfig{figure=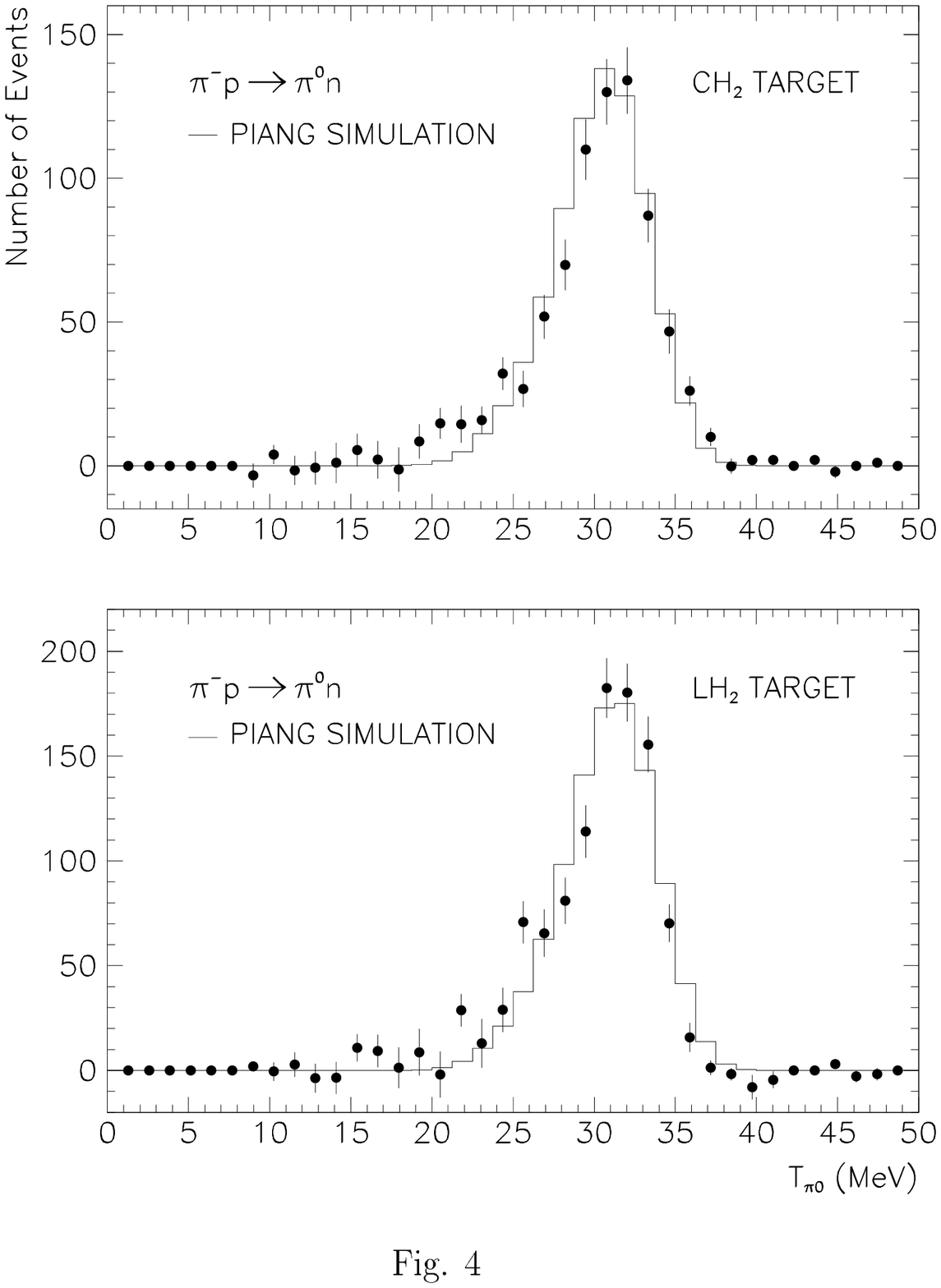}}
\newpage
\centerline{\psfig{figure=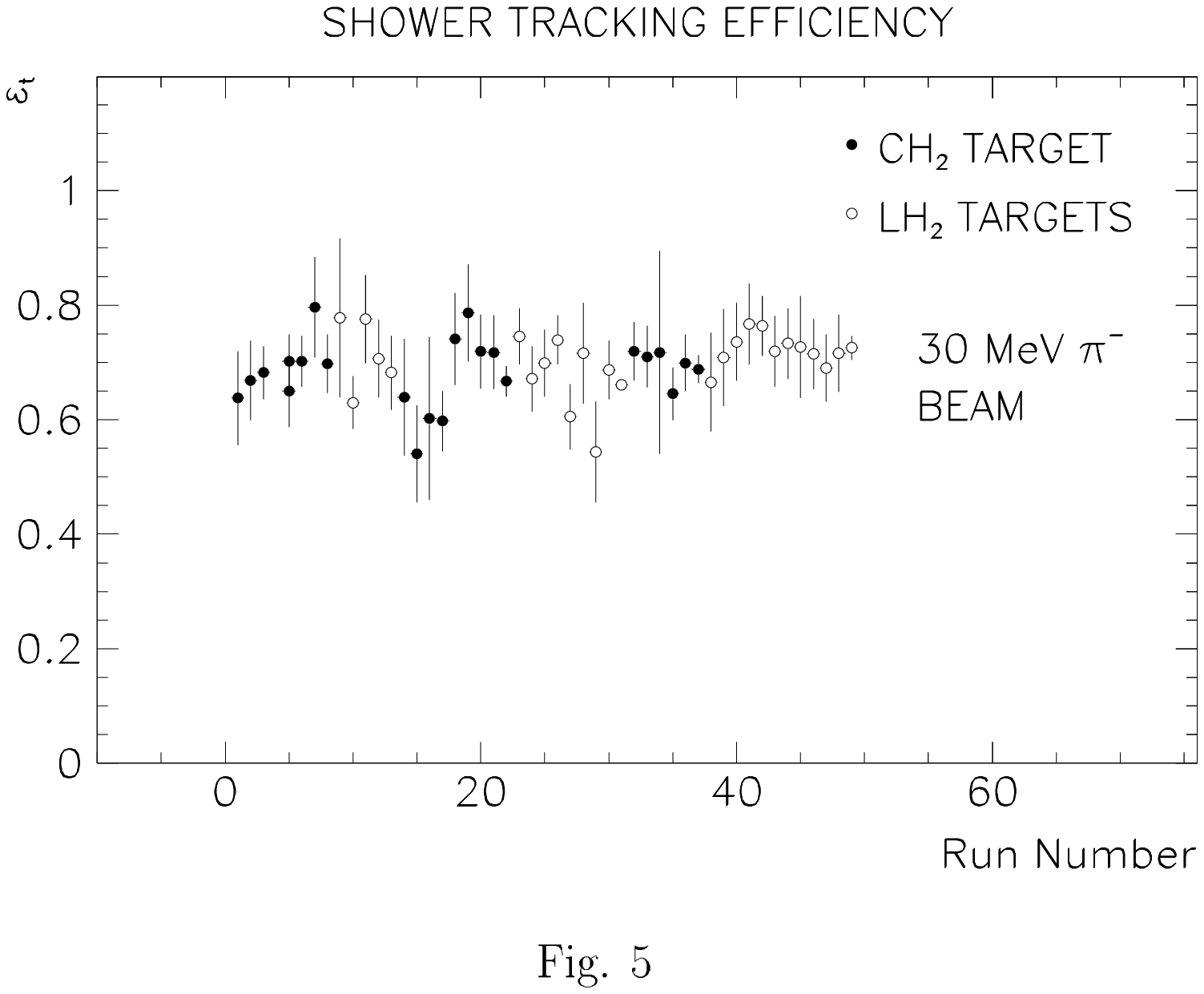}}
\newpage
\vglue -2cm
\centerline{\psfig{figure=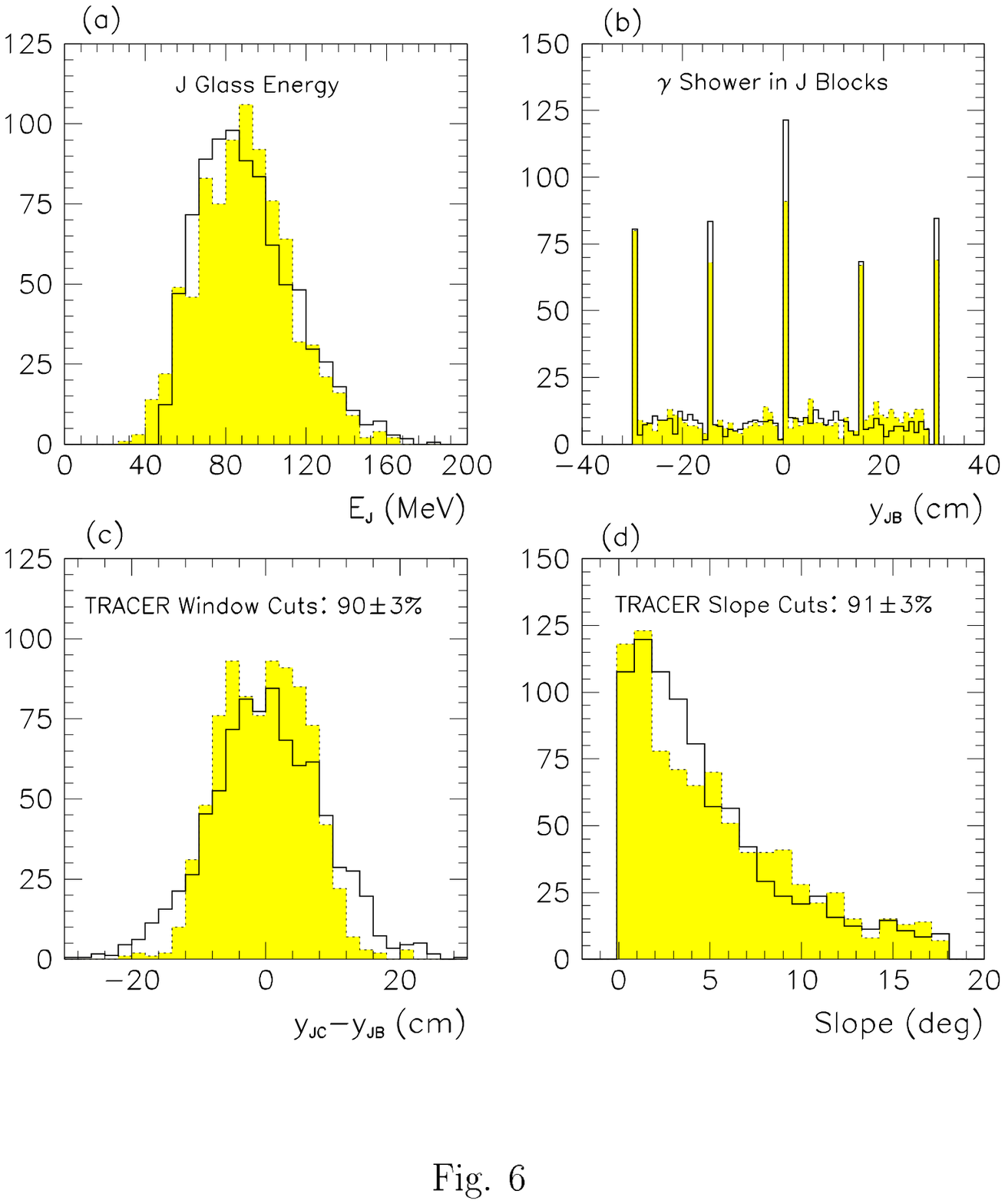}}
\newpage
\centerline{\psfig{figure=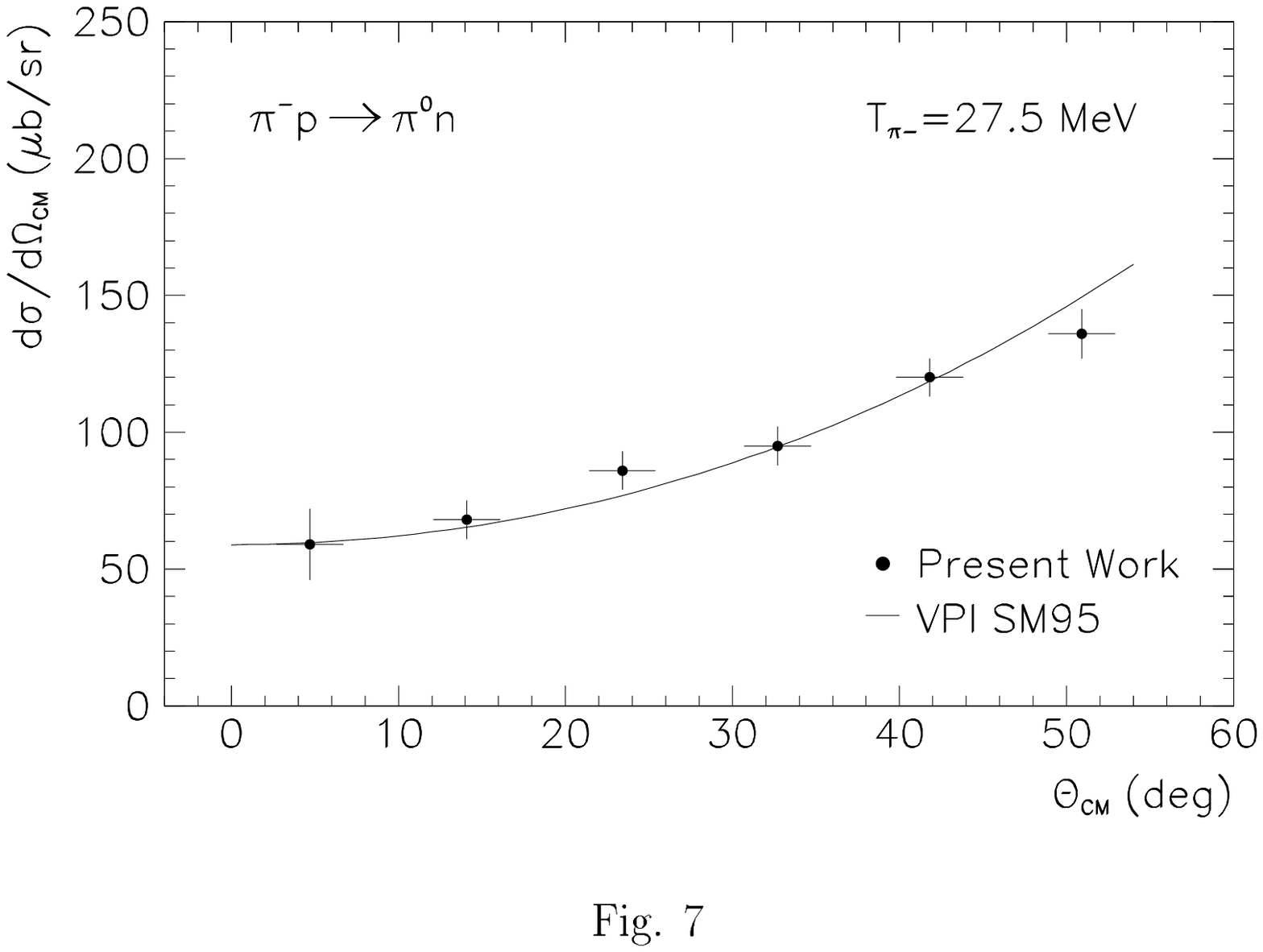}}


\begin{references}

\bibitem[*]{byline} Present address: Physics Department, Hampton University, 
Hampton, VA 23668, USA.
\bibitem[\dag]{byline} Present address:  Thomas Jefferson National 
Accelerator Facility, Newport News, VA 22606, USA.
\bibitem[\ddag]{byline} Present address:  Saskatchewan Accelerator Laboratory,
 University of Saskatchewan,  Saskatoon, Saskatchewan,  CANADA S7N 0W0.
\bibitem[\S]{byline} Present address: Department of Physics, Old 
Dominion University, Norfolk, VA 23529, USA.

\bibitem{arn95} R. A. Arndt, I. I. Strakovsky, R. L. Workman, and
M. M. Pavan, Phys. Rev. C {\bf 52}, 2120 (1995).
\bibitem{arn91} R. A. Arndt, Z. J. Li, L. D. Roper, R. L. Workman, 
and J. M. Ford, Phys. Rev. D {\bf 43}, 2131 (1991).
\bibitem{gib95} W. R. Gibbs, L. Ai, and W. B. Kaufmann, Phys. Rev. Lett.
{\bf 74}, 3740 (1995).
\bibitem{sig86} P. B. Siegel and W. R. Gibbs, Phys. Rev. C {\bf 33},
1407 (1986).
\bibitem{gas82} J. Gasser and H. Leutwyler, Phys. Rep. {\bf 87}, 77 (1982).
\bibitem{koch82} R. Koch, Z. Phys. C {\bf 15}, 161 (1982). 
\bibitem{gas91} J. Gasser, H. Leutwyler, and M. Sainio, Phys. Lett. B
{\bf 253}, 252 (1991).
\bibitem{salom84} M. Salomon, D. F. Measday, and J-M. Poutissou, 
Nucl. Phys. {\bf A414}, 493 (1984).
\bibitem{bagh88} A. Bagheri, K. A. Aniol, F. Entezami, M. D. Hasinoff, 
D. F. Measday, J-M. Poutissou, M. Salomon, and B. C. Robertson, Phys. 
Rev. C {\bf 38}, 885 (1988).
\bibitem{fitzg86} D. H. Fitzgerald, H. W. Baer, J. D. Bowman, 
M. D. Cooper, F. Irom, N. S. P. King, M. J. Leitch, E. Piasetzky,
W. J. Briscoe, M. E. Sadler, K. J. Smith, and J. N. Knudson, Phys. Rev. 
C {\bf 34}, 619 (1986).
\bibitem{sad91} M. E. Sadler, B. M. Brooks, L. D. Isenhower, W. J. 
Briscoe, J. D. Bowman, D. H. Fitzgerald, and J. N. Knudson, in 
{\sl Pion-Nucleon Physics and the Structure of the Nucleon}, Vol. 2, 
p. 131, Bad Honnef, 1991.
\bibitem{Bur75} R. L. Burman, R. L. Fulton, and M. Jakobson,
Nucl. Instr. Methods {\bf 131}, 29 (1975).
\bibitem{But82} G. W. Butler, B. J. Dropesky, C. J. Orth, R. E. L. Green,
R. G. Korteling, and G. K. Y. Lam, Phys. Rev. C {\bf 26}, 
1737 (1982).
\bibitem{poc94} D. Po\v{c}ani\'c, E. Frle\v{z}, K. A. Assamagan,
J. P. Chen, K. J. Keeter, R. M. Marshall, R. C. Minehart, L. C. Smith,
G. E. Dodge, S. S. Hanna, B. H. King, and J. N. Knudson, Phys. 
Rev. Lett. {\bf 72}, 
1156 (1994).
\bibitem{leo87} W. R. Leo, {\sl Techniques for Nuclear and Particle Physics 
Experiments}, (Springer-Verlag, New York, 1987), p. 24, and references 
therein.
\bibitem{leitch91} M. J. Leitch, private communication, LAMPF Experiment
E942, 1992.
\bibitem{kul72} G. Kuhl and U. Kneissl, Nucl. Phys. {\bf A195}, 559
(1972).
\bibitem{ort78} C. J. Orth, M. W. Johnson, J. D. Knight, and
K. Wolfsberg, LAMPF Experiment E331, 1977. 
\bibitem{How87} H. Howard, {\it LAMPF User's Handbook}, MP-DO-1-UHB, 3rd
Ed., LANL, Los Alamos, 1987.
\bibitem{baer81} H. W. Baer, R. D. Bolton, J. D. Bowman, M. D. Cooper,
F. H. Cverna, R. H. Heffner, C. M. Hoffman, N. S. P. King,
J. Piffaretti, J. Alster, A. Doron, S. Gilad, M. A. Moinester,
P. R. Bevington, and E. Winkelmann, Nucl. Inst. Methods {\bf 180},
445 (1981).
\bibitem{brun87} R. Brun,  F. Bruyant, M. Maire, A. C. McPherson, 
and P. Zanarini, {\tt GEANT3}, CERN publication DD/EE/84-1, Geneva, 1987.
\bibitem{hoe82} M. V. Hoehn and D. H. Fitzgerald, LAMPF Report A82-01,
LANL, Los Alamos, 1982.
\bibitem{gilad79} S. Gilad, Ph. D. Thesis, Tel Aviv University, 1979.
\bibitem{gaille84} F. C. Gaille, V. L. Highland, L. B. Auerbach,
W. F. McFarlane, G. E. Hogan, C. M. Hoffman, R. J. Macek,
R. E. Morgado, J. C. Pratt, and R. D. Werbeck, Phys. Rev. D {\bf 30},
2408 (1984).
\bibitem{frlez93} E. Frle\v{z}, Ph. D. Thesis, University of Virginia,
1993; LANL Report LA-12663-T, Los Alamos, 1993.
\bibitem{hubb69} J. H. Hubbel, NSRDS-NBS {\bf 29}, 1969.
\bibitem{baer80} H. W. Baer, {\it The $\pi^0$\/ Spectrometer Notes}, 
LANL, Los Alamos, 1980.
\bibitem{gilad77} S. Gilad, J. D. Bowman, M. D. Cooper, R. H. Heffner,
C. M. Hoffman, M. A. Moinester, J. M. Potter, F. H. Cverna,
H. W. Baer, P. R. Bevington, and M. W. McNaughton, Nucl. Inst. 
Methods {\bf 144}, 103 (1977).
\bibitem{drop79} B. J. Dropesky, G. W. Butler, C. J. Orth, R. A. Williams, 
M. A. Yates-Williams, G. Friedlander, and S. B. Kaufman, Phys. Rev. 
C {\bf 20}, 1844 (1979).
\end{references}
\end{document}